# Production of Tachyonic Neutrino in Matter

Luca Nanni

luca.nanni@student.unife.it

**Abstract** Recently, a new theory based on superluminal tunnelling has been proposed to explain the transition of highly energetic neutrinos propagating in matter to tachyonic states. In this work, we determine the possible mechanisms that lead neutrinos into a superluminal realm based on the assumption that ultrarelativistic neutrinos travelling in matter lose part of their energy with the emission of Bremsstrahlung radiation. The obtained photons, in turn, can create neutrino-antineutrino pairs, one or both of which can be superluminal. We also prove that pair creation may occur with neutrino flavour oscillation provided that only one of them is a space-like particle. This suggests that mass oscillation and superluminal behaviour could be related phenomena. Finally, using the generalised Lorentz transformations, we formulate the Lagrangian of the kinematically allowed scattering processes. The structure of this Lagrangian is consistent with the formalism of the Standard Model. Based on this Lagrangian, at least one of the particles forming the pair must always be subluminal. The possibility that the pair creation process is mediated by a dark photon is also discussed.

*Keywords*: superluminal neutrino, quantum tunnelling, mass oscillation, generalised Lorentz transformations

## 1. Introduction

Neutrino physics is one of the most intriguing, promising fields of research for theoretical physicists [1–4]. It ranges from the physics of high energies tested in particle accelerators [5–6] to astrophysics [7–10], passing through many new theories aiming to explain its anomalies [11–12]. Among these anomalies, that of superluminal neutrinos is the most debated, the publications dedicated to it only multiplying [13–16]. Recently, the controversy about the data of the OPERA experiment in 2011 raised questions about the abilities of neutrinos to travel faster than light. However, the data available from other experiments (Borexino, SciBooNe, Super-K, Minos, Icarus, IceCube, etc.) still leave the possibility of finding the evidence necessary (a single confirmation would be sufficient) to affirm that, besides ordinary matter, there is also tachyon matter. This is likely the reason many physicists study superluminal neutrinos.

As is well-known, the theory of relativity forbids overcoming the speed of light, but, as argued by Sudarshan [21–22], this constraint fails if a massive particle is created from the outset in a superluminal state. The study of tachyons in the framework of quantum mechanics uses models developed in the last few decades that coherently extend the theory of relativity to superluminal motion. Some of these theories violate Lorentz invariance and describe tachyons as unstable particles that decay into ordinary (subluminal) particles [23–24]. Other models instead treat tachyons as stable particles that are parts of the real world [25–26].

The aim of this paper is to complete a recently proposed theoretical model [27] which explains, in a covariant way, the superluminal behaviour of neutrinos in matter. This theory is based on the Hartman effect, in which the tunnelling time of a particle travelling through a potential barrier does not depend on its width (provided that the barrier is wide enough) [28]. Neutrinos propagating in matter interact with baryons and leptons according to the usual scattering mechanisms, passing through some decay processes that we should identify in superluminal states. All these interactions form a potential barrier to superluminal tunnelling. The overall result is particle deceleration in which energy is partially transformed by Bremsstrahlung radiation [29–31]. This process is still being studied, and the mechanism by which it occurs is not yet clear. Millar et al. [32], to study neutrino mass, investigated the process ν+N→N+ν+γ (N is a heavy nucleus) for a neutrino with energy comparable with its mass. Using tungsten (Z=74) as the target, they calculated a cross section that is proportional to $10^{-68} \, cm^2$. Still in the context of neutrino-nucleus interaction, a possible mechanism is the photon emission from weak neutral current interaction [33]. In this case, the estimated cross section is of the order of $10^{-42} \, cm^2$. In turn, the obtained photons can create neutrino-antineutrino pairs in which one or both particles can be in a superluminal state. The mechanism by which the created pairs have different flavours is also discussed. The kinematics of the proposed mechanisms determine which of these are allowed or forbidden. Finally, using the generalised Lorentz transformations (GLTs) proposed by Recami [34–35] and ordinary Dirac fields for neutrinos, we obtain the Lagrangian describing all the kinematically allowed mechanisms. This Lagrangian is invariant under the usual Lorentz transformations, proving that the obtained model describes the tachyonic neutrino within the usual Standard Model formalism. From this Lagrangian, we find that at least one of the neutrinos forming the pair must be subluminal. Finally, the case in which the pair creation is mediated by a dark photon is also discussed.

## 2. Kinematics of Tachyonic Pair Production

In this section, we briefly review some notions of tachyon dynamics to facilitate the study of the kinematics of the tachyonic pair production addressed below.

The reference frame in which an ordinary particle is at rest is characterised by the unitary Lorentz factor $\gamma$ and energy equal to $mc^2$. Things become a bit complicated for tachyons. Considering that, for tachyons, the energy-momentum relationship is $E^2 = p^2c^2 - m^2c^4$, the reference frame where the tachyon is *equivalent* to an ordinary particle at rest is that in which the tachyonic Lorentz factor $\gamma_t$ is unitary [36], to which corresponds a velocity $u_t = \sqrt{2}c$. In this case, the tachyon energy is $E_t^2 = m^2c^4$ and the energy to create a tachyon pair is the same needed to produce a bradyon pair. However, while it is possible in the subluminal case to create pairs in which both particles have null impulses, for tachyon pairs, the speeds of the particles must always be greater than that of light. This means that the production of tachyon pairs is energetically more expensive than the bradyonic case. However, this energy surplus progressively decreases as the speeds of the produced tachyonic pairs increase. Particularly, as the velocity $u_t \to \infty$, the tachyonic energy tends to zero. To this limit, the energy gap between tachyonic particles and antiparticles tends to $2mc^2$ (here, the mass is the absolute value of the imaginary tachyonic value), namely, the same value as in the subluminal case in which pairs are at rest. This suggest that, for tachyons, the *rest* frame is that in which $u_t \to \infty$.

If the Lorentz factor of an ordinary particle and a tachyon of equal mass $m = |m_t|$ has the same value $\gamma = \gamma_t$, and therefore the same energy, the impulse of the superluminal particle is always greater than the bradyonic one:

$$\begin{cases} E_t^2 = \gamma_t^2 m^2 c^4 = p_t^2 c^2 - m^2 c^4 \Rightarrow |p_t| = |m|c(\gamma_t^2 + 1)^{1/2} \\ E^2 = \gamma^2 m^2 c^4 = p^2 c^2 + m^2 c^4 \Rightarrow |p| = mc(\gamma^2 - 1)^{1/2} \end{cases} \quad (1)$$

The two impulses will be equal at the limits $u \to c$ and $u_t \to c$. Finally, we note that, at the limit $u_t \to \infty$, the tachyonic energy goes to zero and the impulse becomes $|p_t| = |m|c$. This means that the Compton wavelength of a tachyon for $u_t \to \infty$ is equal to the Compton wavelength of an ordinary particle in the rest frame. This is further confirmation that the rest frame of a tachyon is that in which the velocity $u_t$ tends to infinity.

## 3. Generalised Lorentz Transformations

In the literature, several works have studied the group of GLTs necessary for coherently extending the theory of relativity to superluminal motions [37–43]. In this paper, we use those proposed by Recami [34–35] to formulate a Lagrangian that is overall invariant under the usual Lorentz transformations despite describing the dynamics of processes involving the production of tachyons. The GLTs are given by the following matrices:

$$\Lambda_{GLT} = \eta \begin{pmatrix} \gamma & i\gamma u_x/c & i\gamma u_y/c & i\gamma u_z/c \\ -i\gamma u_x/c & \delta + \alpha u_x^2/u^2 & \alpha u_x u_y/u^2 & \alpha u_x u_z/u^2 \\ -i\gamma u_y/c & \alpha u_x u_y/u^2 & \delta + \alpha u_y^2/u^2 & \alpha u_y u_z/u^2 \\ -i\gamma u_z/c & \alpha u_x u_z/u^2 & \alpha u_y u_z/u^2 & \delta + \alpha u_z^2/u^2 \end{pmatrix}, \quad (2)$$

where

$$\begin{cases} \delta = \sqrt{(1 - tg^2\theta)/|1 - tg^2\theta|}; \; \gamma = |1 - tg^2\theta|^{-1/2}; \; tg\theta = \beta \\ \text{and } \eta = \delta^2 \frac{\cos\theta}{|\cos\theta|}; \; \alpha = \gamma - \delta. \end{cases} \quad (3)$$

It is easy to verify that, for ordinary particles, $0 \le \theta \le \pi/4$, $\delta = 1$, $\eta = 1$ and $\alpha =$

$\gamma - 1$, while for tachyons $\pi/4 \leq \theta \leq \pi/2$, $\delta = i$, $\eta = -1$ and $\alpha = \gamma - i$. From here on, we set $\Lambda_{LT}$ as the Lorentz transformations and $\Lambda_{SLT}$ as the superluminal Lorentz transformations; both are obtained from matrix (2). For bradyons matrix (2) may be decomposed as the following:

$$\Lambda_{LT} = \gamma \begin{pmatrix} 1 & i\beta_x & i\beta_y & i\beta_z \\ -i\beta_x & 0 & 0 & 0 \\ -i\beta_y & 0 & 0 & 0 \\ -i\beta_z & 0 & 0 & 0 \end{pmatrix} + \begin{pmatrix} 0 & 0 & 0 & 0 \\ 0 & 1 & 0 & 0 \\ 0 & 0 & 1 & 0 \\ 0 & 0 & 0 & 1 \end{pmatrix} + \frac{(\gamma-1)}{\beta^2} \begin{pmatrix} 0 & 0 & 0 & 0 \\ 0 & \beta_x^2 & \beta_x\beta_y & \beta_x\beta_z \\ 0 & \beta_x\beta_y & \beta_y^2 & \beta_y\beta_z \\ 0 & \beta_x\beta_z & \beta_y\beta_z & \beta_z^2 \end{pmatrix}. \quad (4)$$

Therefore, the Lorentz transformation is a linear combination of three Hermitian matrices: first, a pure boost along the time axis while the sum of the second and third matrices is a generic rotation in $\mathfrak{N}^3$. Let us denote these matrices respectively as A, B and C. For the superluminal case, we see that matrix (2) becomes the following:

$$\Lambda_{SLT} = -i\gamma_t A' - iB - i\frac{(\gamma_t-1)}{\beta^2}C', \quad (5)$$

where A' and C' have the same structures of matrices A and C and differ from the latter only by the numerical values of the elements composing them, respectively, being the relativistic tachyonic factor $\beta$ greater than one. In writing matrix (5), the Lorentz factor $\gamma_t$ is calculated by the usual formula $(1 - \beta^2)^{-1/2}$ and, therefore, is an imaginary number. In this way, the GLT matrix can be written regardless of the nature of the particle motion. In fact, comparing matrices (4) and (5), we can write the following general form:

$$\Lambda_{SLT} = f(u)\Lambda_{LT}, \quad (6)$$

where the function $f(u)$ is the following:

$$f(u) = \begin{cases} 1 \text{ when } u < c \\ -i \text{ when } u > c \end{cases}. \quad (7)$$

Denoting the generators of the ordinary Lorentz group as $J_i$ and $K_i$, those of the generalised Lorentz group are $f(u)J_i$ and $f(u)K_i$ and the anticommutation relations become the following:

$$[J_i, J_j] = f(u)\varepsilon_{ijk}J_k; \; [K_i, K_j] = f(u)\varepsilon_{ijk}J_k; \; [J_i, k_j] = -f(u)\varepsilon_{ijk}K_k. \quad (8)$$

We thus obtain the explicit form of the GLT group by which it is possible to transform a given Dirac spinor in any other reference frame, even a superluminal one:

$$\psi \to \Lambda_{GLT}\psi. \quad (9)$$

Since $\psi$ is a Dirac spinor, $\Lambda_{GLT}$ can also be written in the following form:

$$\Lambda_{GLT} = \begin{pmatrix} \Lambda(u) & 0 \\ 0 & \Lambda(-u) \end{pmatrix}, \quad (10)$$

where $\Lambda(\pm u)$ are 2×2 matrices:

$$\begin{cases} \Lambda(u) = exp\{i\boldsymbol{\sigma} \cdot \boldsymbol{\theta}/2 - \boldsymbol{\sigma} \cdot \boldsymbol{\rho}/2\} \\ \Lambda(-u) = [\Lambda^\dagger(u)]^{-1} = exp\{i\boldsymbol{\sigma} \cdot \boldsymbol{\theta}/2 + \boldsymbol{\sigma} \cdot \boldsymbol{\rho}/2\} \end{cases}, \quad (11)$$

where $\boldsymbol{\sigma}$ is a vector whose components are the three Pauli matrices, $\boldsymbol{\theta}$ is the vector whose components are the three angles of rotation about the axis of the reference frame in $\mathfrak{R}^3$, and $\boldsymbol{\rho}$ is a vector whose components are the rapidity of each projection of the velocity. The rapidity is $\rho = arcosh(\gamma)$ and is defined within the range $1 \leq \gamma < \infty$. An ordinary particle $\gamma$ always falls within this range, but a tachyon $\gamma_t$ can take values between zero and infinity. However, if the constraint $u_t \leq \sqrt{2}c$ is set, then the rapidity $\rho$ remains well-defined. Therefore, given a generic $\Lambda_{GLT}$ matrix, it is always possible to find a matrix $U$ that diagonalises it in blocks:

$$U\Lambda_{GLT}U^\dagger = \begin{pmatrix} \Lambda(u) & 0 \\ 0 & \Lambda(-u) \end{pmatrix} \qquad (12)$$

provided that $0 \leq u \leq \sqrt{2}c$.

The GLT may also be written as exponential using generators (8):

$$\Lambda_{GLT} = exp\{-\boldsymbol{\rho} \cdot \boldsymbol{K} - \boldsymbol{\theta} \cdot \boldsymbol{J}\}, \qquad (13)$$

where $\boldsymbol{K}$ and $\boldsymbol{J}$ are vectors whose components are the generators of the GLT group. Comparing eqs. (10) and (13), we find the explicit forms of the 2×2 matrices $\Lambda(\pm u)$:

$$\begin{cases} \Lambda(u) = exp\left\{\begin{bmatrix} (i\theta_z/2 + \rho_z) & (\theta_y + i\theta_x)/2 - (\rho_x - i\rho_y) \\ (-\theta_y + i\theta_x)/2 - (\rho_x + i\rho_y) & -(i\theta_z/2 - \rho_z) \end{bmatrix}\right\} \\ \Lambda(-u) = exp\left\{\begin{bmatrix} (i\theta_z/2 + \rho_z) & (\theta_y + i\theta_x)/2 + (\rho_x - i\rho_y) \\ (-\theta_y + i\theta_x)/2 + (\rho_x + i\rho_y) & -(i\theta_z/2 - \rho_z) \end{bmatrix}\right\} \end{cases} \qquad (14)$$

This result is consistent with that obtained by Recami for a collinear motion along the $x$ axis [42].

## 4. Space-Like Pair Production

As mentioned in the first section, we investigate the possible mechanisms leading to the creation of superluminal neutrinos in matter. The assumption that highly energetic neutrinos propagating in ordinary matter lose energy by Bremsstrahlung radiation with the emission of photons is the starting point of this study. In the literature, several works address this topic from different perspectives. Lobanov and Studenikin investigated a mechanism by which highly energetic massive neutrinos, travelling across ordinary matter, lose energy by photon emission [29]:

$$\nu + fermion \rightarrow \nu + fermion + \gamma. \qquad (15)$$

Mechanism (15) can also lead to neutrinos of different flavours if the incoming neutrinos have intrinsic magnetic dipole momenta. Before the work of Lobanov and Studenikin, other mechanisms, listed below, were also taken into consideration [44–47]:

$$\begin{cases} \nu_i \rightarrow \nu_i + \gamma \ m_{\nu_i} = 0 \text{ with external magnetic field,} \\ \nu_i \rightarrow \nu_i + \gamma \ m_{\nu_i} \neq 0 \text{ with external magnetic field and non} - \text{zero magnetic moment,} \\ \nu_i \rightarrow \nu_j + \gamma \ m_{\nu_i} \neq 0 \text{ with external magnetic field, and} \\ \nu_i \rightarrow \nu_i + \gamma \ m_{\nu_i} \neq 0 \text{ by Cherenkov radiation.} \end{cases} \qquad (16)$$

The first mechanism occurs through vacuum polarisation loops in the presence of external magnetic fields and, as shown by the second mechanism, also becomes feasible for massive neutrinos if they have non-zero magnetic momenta. The third mechanism is a radiative decay that may occur both in matter and in a vacuum. What differentiates mechanism (15) from the ones listed in (16) is the electroweak interaction of the incoming neutrinos with the matter.

In 2006, Lobanov followed his previous work [29] by proposing a new pair-production mechanism of $\nu/\bar{\nu}$ [48] occurring in very dense matter:

$$\gamma \to \nu + \bar{\nu} \text{ in matter.} \tag{17}$$

In this work, we propose a mechanism given by the combination of (15) and (17). Using it, we study the various possibilities that Lobanov did not consider:

$$\begin{cases} \nu_i + fermion \to \nu_i + fermion + \gamma \to \nu_{i_t} + \bar{\nu}_{i_t} \\ \nu_i + fermion \to \nu_i + fermion + \gamma \to \nu_{i_t} + \bar{\nu}_i \text{ or vice versa,} \\ \nu_i + fermion \to \nu_i + fermion + \gamma \to \nu_{j_t} + \bar{\nu}_{i_t} \text{ or vice versa, and} \\ \nu_i + fermion \to \nu_i + fermion + \gamma \to \nu_{j_t} + \bar{\nu}_i \text{ or vice versa.} \end{cases} \tag{18}$$

Therefore, we study the case in which the particles of the pair are both tachyons and the case in which only one of them is a tachyon, and for these two mentioned cases, we verify if the mechanism by which one particle changes flavour is kinematically feasible. A mechanism like those proposed in (18) also occurs in the case of an ordinary particle [49]:

$$e^-Z \to e^-Z + \gamma \to \mu^+ + \mu^- \text{ in matter} \tag{19}$$

where $Z$ is a nuclear target.

Let us start with the first mechanism (18). As is well-known, the pair creation $e^+/e^-$ from a photon occurs only in the presence of an external electric field, like that of an atomic nucleus. Otherwise, the conservation of relativistic energy and impulse do not hold. Things change, however, if both particles are tachyons. In fact, using eq. (1), the conservation of energy and impulse can be written as the following:

$$\begin{cases} h\nu = 2\gamma_t m_t c^2 = 2(p_t^2 c^2 - |m_t|^2 c^2)^{1/2} \\ \frac{h\nu}{c} = p_t(\nu)\cos\theta + p_t(\bar{\nu})\cos\theta = 2\gamma_t m_t u_t \cos\theta \end{cases}. \tag{20}$$

We note that $m_t$ and $\gamma_t$ are both imaginary and that their product is both real and positive. Moreover, as usually expected in pair creation, the module of the vector impulse is the same for both particles $\nu$ and $\bar{\nu}$. From the second equation of (20), we obtain the following:

$$h\nu = 2\gamma_t m_t u_t c(\cos\theta) = 2\gamma_t m_t c^2 \beta \cos\theta. \tag{21}$$

Since $\beta > 1$ and $\cos\theta \leq 1$, we can always find a superluminal velocity $u_t$ and an angle $\theta$ for which the product $\beta\cos\theta$ is unitary so that eq. (21) is equal to the first equation of (20). Therefore, from a photon, it is always possible to create a $\nu/\bar{\nu}$ tachyon pair, even in vacuum, due to the fact that, in the momentum-energy dispersion relation of a tachyon, the mass term is always negative. From eq. (21), we also obtain the scattering angle once we set the velocity $u_t$:

$$\beta \cos\theta = 1 \Rightarrow \theta = arcos(1/\beta). \tag{22}$$

The superluminal velocity $u_t$ is obtained solving the first equation of (20):

$$u_t = \pm \frac{(h^2\nu^2 + 4|m_t|^2 c^4)^{1/2}}{2\gamma_t m_t c} \tag{23}$$

Substituting eq. (23) into eq. (22), we obtain the following:

$$\theta = arcos[\pm 2\gamma_t m_t c^2/(h^2\nu^2 + 4|m_t|^2 c^4)^{1/2}] \tag{24}$$

It is observed that, as $\gamma_t \to 0$, that is, as the velocity $u_t$ goes to infinity, the scattering angle tends to $\pm\pi/2$. This, however, is not an acceptable condition, as it would violate the conservation of the impulse written in vector form. The argument of $arcos$ function must range between -1 and 1. This constraint limits the absolute value of the upper tachyon velocity:

$$|2\gamma_t m_t c^2/(h^2\nu^2 + 4|m_t|^2 c^4)^{1/2}| \leq 1 \Rightarrow u_t \leq c[1 + 4|m_t|^2 c^4/(h^2\nu^2 + 4|m_t|^2 c^4)]^{1/2}. \tag{25}$$

Since $h\nu$ is always positive, it follows that $u_t \leq \sqrt{2}c$, to which corresponds a tachyonic Lorentz factor ranging between one and infinity, just like the subluminal case. We thus prove that the creation of a $\nu/\bar{\nu}$ tachyon pair occurs both in matter and a vacuum but the velocity, and so the total energy, has an upper bound. At this limit, it is easy to verify that the energy of photons is $h\nu = 2|m_t|c^2$, as in the case of the pair creation of ordinary particles. As $u_t \to c$, photon energy progressively increases.

Let us now consider the second mechanism (18) by which a pair is created by a tachyon and an antibradyon (or vice versa) with the same masses and energies. This means that the two Lorentz factors $\gamma$ and $\gamma_t$ must be equal and, consequently, that in this scattering mechanism the tachyon energy will also have an upper bound at $m_t c^2$ (at which corresponds the velocity $u_t = \sqrt{2}c$). Following the same approach adopted for the kinematic study of the previous scattering mechanism, we obtain the following:

$$h\nu = mc\left[\gamma u + \sqrt{\gamma^2 u^2 + 2c^2}\right]\cos\theta = mc(\gamma c + |\gamma_t|c) = 2\gamma mc^2. \tag{26}$$

To simultaneously comply with both energy and impulse conservation, one must have the following:

$$\left[\gamma u + \sqrt{\gamma^2 u^2 + 2c^2}\right]\cos\theta = 2\gamma c. \tag{27}$$

From eq. (27), we see that, when $\gamma \to 1$, the cosine of the scattering angle is greater than one, which is not acceptable. Always using eq. (27), if we study the condition under which $0 \leq \cos\theta \leq 1$, we obtain a second-order polynomial equation in $u$ whose discriminant is imaginary. Therefore, the mechanism leading to a neutrino pair—one tachyon and the other bradyon—is not kinematically possible.

The scattering mechanism by which both neutrinos are superluminal but with different imaginary masses must now be investigated. Let us denote by $m_t$ and $m'_t$ the imaginary masses of the two tachyons (it is not necessary to distinguish which of them is the tachyon and which the antitachyon). The energies of the two particles must be equal and, since they have different mass, this is possible if the two impulses are different. Under these hypotheses, eq. (20) becomes the following:

$$\begin{cases} h\nu = (p_t^2 c^2 - |m_t|^2 c^2)^{1/2} + (p'^2_t c^2 - |m'_t|^2 c^2)^{1/2} \\ \frac{h\nu}{c} = p_t \cos\theta + p'_t \cos\theta = \gamma_t m_t u_t \cos\theta + \gamma'_t m'_t u'_t \cos\theta \end{cases}. \quad (28)$$

By straightforward steps, we obtain the following:

$$\cos\theta = \left[\frac{c^2(|m'_t|^2 - |m_t|^2)}{(\gamma' m'_t u'_t)^2 - (\gamma m_t u_t)^2}\right]^{1/2}. \quad (29)$$

If we set $m_t = m'_t$ and $u_t = u'_t$, we obtain $\cos\theta < 1$, confirming the result obtained for the first of the scattering mechanisms (18). If instead, for instance, we set $m_t < m'_t$, then $u_t < u'_t$ in order to obtain a positive value under the square root. Moreover, to ensure that $\cos\theta < 1$, the following inequality must hold:

$$\frac{u'^2}{u'^2 - c^2} > \frac{u^2}{u^2 - c^2}, \quad (30)$$

which can never be verified if $u_t < u'_t$. We thus obtain a contradiction which proves the impossibility of creating a tachyon-antitachyon pair with different masses.

Things change if one of the two particles of the pair is a bradyon (the last scattering mechanism to investigate). In this case, eq. (28) is the following:

$$\begin{cases} h\nu = (p^2 c^2 + m^2 c^2)^{1/2} + (p_t^2 c^2 - |m_t|^2 c^2)^{1/2} \\ \frac{h\nu}{c} = p\cos\theta + p_t \cos\theta = \gamma m u \cos\theta + \gamma_t m_t u_t \cos\theta \end{cases}. \quad (31)$$

As usual, the two created particles must have the same energy:

$$(p^2 c^2 + m^2 c^2)^{1/2} = (p_t^2 c^2 - |m_t|^2 c^2)^{1/2} \Rightarrow p_t^2 = p^2 + (m^2 + |m_t|^2)c^2. \quad (32)$$

By simple calculations, we obtain the following:

$$\cos\theta = \frac{2c\sqrt{\gamma^2 m^2 u^2 + m^2 c^2}}{\left[\gamma m u c + c\sqrt{\gamma^2 m^2 u^2 + (m^2 + |m_t|^2)c^2}\right]}. \quad (33)$$

It is easy to prove that $0 \leq \cos\theta \leq 1$ if and only if $|m_t| \geq 3m$. We conclude that a pair creation in which both particles have the same energies but different flavours is always possible if one of them is a tachyon and the other a bradyon under the condition that the absolute value of the tachyonic mass is greater than three times the bradyonic one. This result suggests that neutrino oscillation and superluminality are related phenomena.

**5. Lagrangian of Oscillating Superluminal Neutrino**

The formulation of the Lagrangian field for the two kinetically possible mechanisms encounters various difficulties. First, we must find the kinetic component of the tachyonic field, which violates the Lorentz invariance. Then, we must introduce an interaction term between the tachyonic and bradyonic neutrino fields with the electromagnetic one. All these terms must be Lorentz-invariant. Our approach is to transform all the terms that violate the Lorentz invariance through the GLT matrices so that they comply with the formalism of the Standard Model.

To solve the first difficulty, we use the Tanaka Lagrangian [50], which is a pseudoscalar for which $p_\mu p^\mu = -m_t^2$ that leads to the following equation of motion:

$$(i\partial_\mu \gamma^5 \gamma^\mu - m_t)\psi_t = 0. \tag{34}$$

Eq. (34) is the Dirac equation for a tachyonic neutrino and differs from the ordinary equation with the presences of the $\gamma^5$ matrix and the real mass term. The Tanaka Lagrangian was used by Chodos to describe tachyonic neutrinos [51]. Its explicit form is the following:

$$\mathcal{L}_t = i\bar\psi_t \gamma^5 \gamma^\mu \partial_\mu \psi_t - m_t \bar\psi_t \psi_t = 0 \tag{35}$$

where $\bar\psi_t = \psi_t^\dagger \gamma^0$, $\gamma^5 = i\gamma^0 \gamma^1 \gamma^2 \gamma^3$, $(\gamma^5)^2 = \mathbb{1}$ and $\hbar = c = 1$. We must notice that matrix $\gamma^5$ is not commutative and allows obtaining from eq. (34) the Klein-Gordon equation for a tachyon:

$$(i\partial_\mu \gamma^5 \gamma^\mu - m_t)(-i\partial_\mu \gamma^5 \gamma^\mu + m_t)\psi_t = (\partial_\mu \partial^\mu - m_t^2)\psi_t = 0. \tag{36}$$

Now we must represent the tachyonic field by means of the bradyonic one; this is the necessary condition to obtain a homogeneous total Lagrangian where only the Dirac neutrino appears. To do this, it is sufficient to transform the Dirac field of the bradyonic neutrino using a GLT matrix:

$$\Lambda_{GLT}\psi \rightarrow \psi_t. \tag{37}$$

Substituting eq. (37) into eq. (35), we obtain the following:

$$\mathcal{L}_t = i(\Lambda_{GLT}\psi)^\dagger \gamma^0 \gamma^5 \gamma^\mu \partial_\mu (\Lambda_{GLT}\psi) - |m_t|(\Lambda_{GLT}\psi)^\dagger \gamma^0 (\Lambda_{GLT}\psi) = 0. \tag{38}$$

Using eqs. (6) and (7) and some Hermitian algebra, we obtain the following:

$$\mathcal{L}_t = i\bar\psi[\Lambda^\dagger \gamma^5 \gamma^\mu \Lambda \partial_\mu]\psi - |m_t|\bar\psi[\Lambda^\dagger \Lambda]\psi = 0 \tag{39}$$

where $\Lambda = -i\Lambda_{LT}$. It is easy to see that the Lagrangian eq. (39) is Lorentz-invariant and can therefore be handled by the formalism of Standard Model.

We must now find the interaction term of tachyonic neutrinos with the electromagnetic field. For this, we recall the Lagrangian of quantum electrodynamics given by the following:

$$\mathcal{L}_{int.} = ig\bar\psi[A_\mu \gamma^\mu]\psi, \tag{40}$$

and we change the covariant derivative to comply with the gauge invariance $U(1)$ for the Chodos equation. We thus obtain the tachyonic interaction term:

$$\mathcal{L}_{int.}^t = ig\bar\psi_t[A_\mu \gamma^5 \gamma^\mu]\psi_t \tag{41}$$

where the potential vector $A_\mu$ can be written as the linear combination of the non-matrix bosons $W_\mu^\pm$ and $B_\mu$ [52]:

$$A_\mu = W_{11\mu}sin\theta_w + B_\mu cos\theta_w. \tag{42}$$

Substituting transformation (37) into eq. (41), we obtain the following:

$$\mathcal{L}_{int.}^{t} = ig\bar{\psi}[\Lambda^{\dagger}(A_{\mu}\gamma^{5}\gamma^{\mu})\Lambda]\psi. \tag{43}$$

We now have all that is needed to write the total Lagrangian for the kinematically accepted mechanisms:

$$\mathcal{L}_{tot} = -\frac{1}{4}F_{\mu\nu}F^{\mu\nu} + \{i\bar{\psi}\Lambda^{\dagger}[\gamma^{5}\gamma^{\mu}(\partial_{\mu} + igA_{\mu}) - |m_{t}|]\Lambda\psi\} \\ + \{i\overline{\psi'}\Lambda^{\dagger}[\gamma^{\mu}(\partial_{\mu} + ig'A_{\mu}) - m]\Lambda\psi'\} \tag{44}$$

where $\psi$ and $\psi'$ are the fields of neutrinos with different masses. This Lagrangian is Lorentz-invariant and complies with the constraints $c \leq u_{t} \leq \sqrt{2}c$ and $|m_{t}| \geq 3m$. Concerning the gauge-coupling parameters $g$ and $g'$, if supposing that, in principle, they are not the same, that is, that the interaction of neutrinos and photons does not depend on their tachyonic or bradyonic natures, we cannot say anything. Without some experimental evidence, however, we risk of entering an excessively speculative ambit.

## 6. Discussion

This study is inspired by the ideas of other theoretical works [44–47] according to which high-energy neutrinos in matter can interact with their constituents (leptons and baryons), losing energy by radiative emission. The photons, in turn, can create neutrino-antineutrino pairs [29,49]. One might expect that the interaction of neutrinos and photons is weak and constitutes mainly an astrophysical interest. The standard model does not predict coupling at the tree level between neutrinos and photons. However, as investigated by Karl and Novikov [53], the neutrino-photon interaction may occur through loops, where virtual leptons couple both weakly and electromagnetically. Therefore, we argue that this interaction can be considered a reasonable starting point for this work, the purpose of which is to extend this result to superluminal neutrinos.

First, we investigated the kinematics of the processes that lead to the creation of pairs in which one or both particles are in tachyonic states. The conservation of energy and momentum holds only for the mechanism that leads to the formation of tachyonic pairs, and for that, only when one of the two particles is in a superluminal state with a different mass. In the first case, the velocities of the two tachyons are upper bounded, while for the second mechanism, the absolute value of the tachyon mass must be at least three times that of the bradyon. The constraint to which the first mechanism is subject proves that the tachyonic pair is unstable and will tend to decay according to a mechanism already studied by Jentschura [15,16]. The constraint on the second mechanism, on the other hand, is consistent with the current hypotheses on the masses of the three known neutrinos.

Second, we formulated the Lagrangian that describes both the kinematically possible processes, making use of the Dirac field and the GLT transformations. The Lagrangian thus obtained is Lorentz-invariant. The four-current calculated by this Lagrangian is positive definite only if at least one of the particles of the pair is a bradyon. This proves that the neutrino oscillation is mediated by a superluminal state: the tachyonic particle is created with a different mass than the initial neutrino and then decays into an ordinary neutrino (always according to the mechanism proposed by Jentschura). This result confirms what has already been conjectured by other authors, who believe that the neutrino mass oscillation occurs at tachyonic group velocities [54–56].

In a more speculative framework, we can assume that the Bremsstrahlung process leads to the formation of a massive dark photon. Recently, this hypothetical particle, mediator of a force not contemplated by the standard model, has attracted the attention of theoretical physicists since, weakly interacting with both ordinary and dark matter, it could explain some experimental phenomena to which physics is not still able to answer (for instance, neutrino oscillation and baryon asymmetry) [57-59]. In such a case, the first term of the Lagrangian (44) must be replaced by a new one given by the following:

$$\mathcal{L}_{D.P.} = \frac{1}{2} m_{D.P.}^2 A'_\mu A'^\mu - \frac{1}{4} F'_{\mu\nu} F'^{\mu\nu} - \frac{\varepsilon}{2} F'_{\mu\nu} F^{\mu\nu} \tag{45}$$

where $A'_\mu$ is the dark photon field, $m_{D.P.}$ is its mass and $\varepsilon$ is the kinetic mixing parameter that allows the dark photon to couple to ordinary matter [60]. In the total Lagrangian (44), we have then replaced the field $A_\mu$ with $A'_\mu$; it is also expected that both the two coupling parameters $g$ and $g'$ are different from the previous case. In conclusion, in the case of the creation of the $\nu_t/\bar{\nu}$ ($\nu/\bar{\nu}_t$) pair, the process is mediated by loops in which the photons produced by the Bremsstrahlung process interact with virtual charged particles. If the produced photons are dark, then the process is mediated by $U'(1)$ gauge bosons. $U'(1)$ is a new Abelian group the symmetry of which is spontaneously broken to give mass to dark bosons. This last may weakly interact both with ordinary and dark matter through the kinetic parameter $\varepsilon$. If neutrinos really interact with dark photons, then it is possible that the latter may decay into $\nu/\bar{\nu}$ pairs both ordinary and tachyonic. Dark bosons thus behave similarly to $Z$ bosons, which decay into lepton-antilepton pairs, including neutrino-antineutrino ones.

The theory proposed in this study could also help to interpret some experimental evidence compatible with a neutrino with a negative square mass. I refer to the MINOS experiment [61] and to those relating to the determination of the mass of charged pions [62-64]. Furthermore, this theory is proposed as an alternative to other theories that explain the superluminal behaviour of the neutrino because of its interaction with curved spacetime [65], even if they have as a common point the sensitivity of the neutrino to the environment. The innovative element of the present theory remains the connection between flavour oscillation and superluminal behaviour, a feature that differentiates it from the others available in the recent literature on tachyon neutrinos.